\documentclass{article}

\usepackage{PRIMEarxiv}
\pdfoutput=1
\usepackage[utf8]{inputenc} 
\usepackage[T1]{fontenc}    
\usepackage{hyperref}       
\usepackage{url}            
\usepackage{booktabs}       
\usepackage{amsfonts}       
\usepackage{nicefrac}       
\usepackage{microtype}      
\usepackage{lipsum}
\usepackage{fancyhdr}       
\usepackage{graphicx}       
\graphicspath{{media/}}     
\usepackage{multirow}
\usepackage{tabularx}
\usepackage{caption}
\usepackage{longtable}
\usepackage{graphicx}
\usepackage[numbers]{natbib}
\usepackage{float} 
\fancypagestyle{firstpage}{
    \fancyhf{} 
    \fancyfoot[C]{%
        \begin{flushleft}
            \hrulefill\\
            {Corresponding author: kelkaeazle@swinburne.edu.my}
        \end{flushleft}
    }
}

\title{RetSeg: Retention-based Network for Polyps Segmentation}
\author{
    \begin{tabular}{p{0.4\textwidth}@{\hspace{2em}}p{0.4\textwidth}}
    Khaled ELKarazle & Valliappan Raman \\
    \texttt{kelkaeazle@swinburne.edu.my} & \texttt{valliappan@cit.edu.in} \\
    Swinburne Univ. of Technology (Malaysia) & Coimbatore Institute of Technology (India) \\
    \\
    Caslon Chua & Patrick Then \\
    \texttt{cchua@swinburne.edu.au} & \texttt{pthen@swinburne.edu.my} \\
    Swinburne Univ. of Technology (Australia) & Swinburne Univ. of Technology (Malaysia) \\
    \end{tabular}
}

\begin{document}
\maketitle
\thispagestyle{firstpage}
\begin{abstract}
Vision Transformers (ViTs) have revolutionized medical imaging analysis, showcasing superior efficacy compared to conventional Convolutional Neural Networks (CNNs) in vital tasks such as polyp classification, detection, and segmentation. Leveraging attention mechanisms to focus on specific image regions, ViTs exhibit contextual awareness in processing visual data, culminating in robust and precise predictions, even for intricate medical images. Moreover, the inherent self-attention mechanism in Transformers accommodates varying input sizes and resolutions, granting an unprecedented flexibility absent in traditional CNNs. However, Transformers grapple with challenges like excessive memory usage and limited training parallelism due to self-attention, rendering them impractical for real-time disease detection on resource-constrained devices. In this study, we address these hurdles by investigating the integration of the recently introduced retention mechanism into polyp segmentation, introducing RetSeg, an encoder-decoder network featuring multi-head retention blocks. Drawing inspiration from Retentive Networks (RetNet), RetSeg is designed to bridge the gap between precise polyp segmentation and resource utilization, particularly tailored for colonoscopy images. We train and validate RetSeg for polyp segmentation employing two publicly available datasets: Kvasir-SEG and CVC-ClinicDB. Additionally, we showcase RetSeg's promising performance across diverse public datasets, including CVC-ColonDB, ETIS-LaribPolypDB, CVC-300, and BKAI-IGH NeoPolyp. While our work represents an early-stage exploration, further in-depth studies are imperative to advance these promising findings.
\end{abstract}


\section{Introduction}
Transformers [1] made their debut in the field of natural language processing, revolutionizing the construction of large language models. They addressed several challenges encountered by previous recurrent models such as handling long-term dependencies, dealing with training time inefficiencies due to the sequential nature of RNNs, understanding context especially with homonyms and synonyms, handling ambiguity and idiomatic phrases, and adapting to language change and multilingual data. Their introduction marked a significant advancement in the field [2]–[4]. Shortly after the advent of Transformers in NLP tasks, computer vision researchers began to incorporate Vision Transformers [5] into various tasks, including classification, object detection, and semantic segmentation. Vision Transformers (ViTs) have pioneered advancements in computer vision by harnessing the power of self-attention mechanisms and the ability to capture long-range dependencies [6], [7]. Unlike traditional Convolutions Neural Networks (CNNs), ViTs process visual data in a context-aware manner, leading to more accurate and robust predictions. The ability to understand the broader context of an image beyond local features has marked a significant leap in tasks such as image classification, object detection, and semantic segmentation [8]. In the domain of medical imaging, particularly in the context of colonoscopy, precise segmentation of medical images is essential for early disease detection and effective treatment planning.

Accurate detection of regions of interest, such as polyps, is fundamental for precise diagnosis and timely interventions. However, existing medical image segmentation methods encounter challenges in accurately capturing intricate textures and refining boundaries due to the lack of global context. 

ViTs have emerged as a transformative solution to improve the detection and classification accuracies in medical imaging, by effectively capturing complex and challenging textures present in various types of medical images. Leveraging self-attention mechanisms, ViTs process intricate textures in a context-aware manner, leading to more accurate and robust predictions [9] - [11]. Despite the impressive performance demonstrated by Vision Transformers, they encounter several challenges, particularly when applied to tasks such as medical image segmentation. These challenges include a lack of spatial and local context to refine organ boundaries, and a limited ability to generalize well on small medical imaging datasets, which often leads to a reliance on large-scale pre-training [12], [13].   Moreover, in the context of NLP, the self-attention mechanism, while powerful, is not without its own set of challenges. These include difficulties in modeling periodic finite-state languages and hierarchical structures, dependence on a single input symbol for language acceptance, and struggles with handling very long sequences due to their quadratic complexity [14], [15]. 

On the other hand, from the perspective of Vision Transformers (ViTs), these challenges extend to a quadratic computational overhead for high-resolution vision tasks, sparsity in self-attention during inference, a lack of inductive biases leading to the need for large volumes of training data or increased model capacity, interpretability challenges due to the fundamental difference between how multi-head self-attention works in image and language settings, and issues with perceptual grouping where studies suggest that the attention mechanisms in vision transformers perform similarity grouping and not attention [16]–[20]. 
Considering these challenges, the deployment of transformer-based models for medical image segmentation on low-end devices for real-time detection continues to be a significant hurdle in both the medical and deep learning fields.

In response to the challenges of substantial computational overhead and the absence of training parallelism, while preserving satisfactory accuracy, Sun et al. 2023 [21] introduced the concept of retention through the development of Retentive Networks. The fundamental difference between retention and attention is the incorporation of decay masks. These masks are utilized to control the attention weights of each token in relation to its following tokens. Moreover, in contrast to attention, retention processes a sequence of tokens instead of processing them individually which leads to efficient usage of computational resources. 

This mechanism enhances the model’s ability to effectively capture prior knowledge, leading to improved performance. Retentive Networks operate under several paradigms, including Parallel Representation, Recurrent Representation, and Chunkwise Recurrent Representation.
Thus far, the concept of retention has not been as extensively explored in computer vision tasks as Vision Transformers. Only a single study by Fan et al. 2023 [22] has pioneered the introduction of Retentive Networks into the field of computer vision, presenting the RetNet and Transformer Concept (RMT). Inspired by the results and findings of Fan et al., we develop interest into the prospect of incorporating retention into medical image segmentation, with a particular focus on polyp segmentation. 

We introduce RetSeg, a binary segmentation network modeled after U-NET [23], which employs Multi-Head Retention blocks specifically designed for polyp segmentation. Partially inspired by the architectural principles of TransUNET [24] and TransResU-Net [25], we incorporate a retention mechanism into the bottleneck module of our network. RetSeg is composed of three integral components: 1) The Encoder, which is responsible for feature extraction, 2) The Retention Block, which applies the retention mechanism to capture complex dependencies, and 3) The Decoder, which reconstructs the segmented output from the encoded features.

For the retention mechanism, we only follow the parallel representation paradigm; however, more experiments will be conducted to study the possibility of introducing both Chunkwise Recurrent Representation and the Recurrent Representation paradigms which were originally introduced in Retentive Networks. Details on each of the components are discussed in the subsequent sections. We use a combination of loss functions namely binary cross-entropy, Dice loss, focal loss, and L1 loss to optimize our network.

We train RetSeg on a combined dataset of Kvasir-SEG [26] and CVC-Clinic [27]. We rigorously evaluate RetSeg on four distinct public colonoscopy datasets, namely: CVC-300 [73], CVC-ColonDB [74], ETIS-LaribPolypDB [75] and BKAI-IGH NeoPolyp [76]. We monitor metrics such as Intersection-over-Union (IoU), Dice Coefficient, Mean Squared Error, Precision, Recall and F1. We conduct a comparative study with existing polyps’ segmentation methods. Our contributions can be summarized in the following points:
\\\\\\\\
\begin{itemize}
\item We present RetSeg, a novel encoder-decoder network that leverages a retention mechanism for the segmentation of polyps.

\item We propose the use of Multi-Head Retention blocks, a unique approach to apply retention to tokenized image patches.

\item	We perform comprehensive experiments on four distinct datasets and carry out a comparative study to elucidate the differences and advantages of RetSeg over existing polyp segmentation methods.

\end{itemize}

The rest of this paper is organized into sections covering background \& related work, our proposed method, results, discussion, and conclusion.

\section{Background \& Related work}
\subsection{Colorectal Polyps}
Colorectal polyps, arising from the mucosal surface of the colon or rectum due to uncontrolled cell replication, manifest as abnormal growths. These growths may be neoplastic or non-neoplastic, necessitating histopathology for confirmation. Colorectal polyps left undetected and untreated can progress into colorectal cancer over time [28]. Several factors, such as lifestyle, diet, and genetic predisposition, have been identified as contributors to their development. Modifiable risk elements encompass smoking, obesity, high red meat consumption, and low fiber and calcium intake [29], [30].

The gold standard for detecting and treating colorectal polyps is colonoscopy, a multistep visual examination procedure. Bowel preparation, involving a bowel-cleansing agent and a special diet for fecal matter clearance, is mandatory prior to the procedure [31], [32]. During colonoscopy, a physician utilizes a colonoscope—a flexible tube with a light source and camera—to traverse the colon by inserting it through the patient's rectum. Specialized instruments passed through the colonoscope's working channel facilitate the removal of polyps and abnormal tissues. Additionally, tissue samples, known as biopsies, can be obtained for histopathological examination.

Although colonoscopy is regarded as the gold standard for colorectal polyp detection, it is not without flaws. Reported miss rates range from 6\% to 28\% [33], [34] Factors contributing to missed polyps include poor bowel preparation, suboptimal visualization of specific colon regions, and low contrast between flat, small polyps and the surrounding mucosa. Endoscopist-related factors, including training, experience, and fatigue, can also influence the miss rate [35].  Due to the heightened risk of colorectal cancer associated with missed polyps, recent years have seen the proposal of deep learning methods as an adjunct to colonoscopy for their detection. Deep learning algorithms, specifically semantic segmentation algorithms, have proven to be promising in enhancing the detection of small and flat polyps which are often overlooked during conventional colonoscopy.

\subsection{Automatic Polyps Detection}
Polyp segmentation is generally viewed as a supervised learning task. In this task, a binary segmentation model is trained using an input image and a corresponding ground truth mask, which highlights the location of the polyp in the image. Over multiple training epochs, the model attempts to accurately predict a binary mask that corresponds the polyp’s location in a given image. This iterative process allows the model to improve its segmentation accuracy over time [36]. 

Polyp segmentation is generally a challenging task due to the complex nature of the input images. These images may be affected by various factors such as poor bowel preparation, the presence of flat and serrated polyps that blend into the background, and limitations in the colonoscope’s field of view, among others. These issues add layers of complexity to the task with many models struggling to precisely segment polyps [37]–[39]. Vision Transformers were initially proposed as a solution to capture long-range dependencies in complex samples. However, Transformers on their own are not adept at capturing rich feature maps that encompass both global and local information. 

To address the issue of insufficient local context, several studies have introduced hybrid approaches that combine the strengths of both Transformers and traditional Convolutional Neural Networks (CNNs). These hybrid models aim to leverage the global receptive field of Transformers while preserving the local feature extraction capabilities of CNNs, thereby providing a more comprehensive understanding of the input data. This combination facilitates the extraction of rich feature maps that encapsulate both local features and long-range dependencies. As a result, it leads to the creation of more precise segmentation masks [40]–[43].

In a recent study by [44] the authors introduced RT-Net, a transformer-based network designed for polyp segmentation in colonoscopy images. The network employs a pyramid vision transformer as the encoder to capture long-range dependencies and global context. It also incorporates three modules: 1) A residual multiscale module, 2) A Region-enhanced attention module, and 3) A feature aggregation module. These modules were designed to learn multi-scale features, enhance polyp regions, and fuse features efficiently. The network was tested on five benchmark polyp datasets: Kvasir-SEG, EndoScene, CVC-ClinicDB, ETIS-LaribPolypDB, and CVC-ColonDB.

The authors in [45] proposed a method known as META-Unet. This approach leverages a Multi-Scale Efficient Transformer Attention (META) mechanism, designed to capture both long-term dependencies and fine-grained features, facilitating adaptive feature fusion. The META mechanism is incorporated into a U-shaped encoder-decoder architecture, termed META-Unet, which is capable of achieving rapid and highly accurate polyp segmentation. The effectiveness of this method was evaluated using four public polyp segmentation datasets: CVC-ClinicDB, Endoscenestill, Kvasir-SEG, and ETIS- LaribPolypDB.

In a different study by [46] the authors introduce a polyp segmentation network that combines CNN and Transformers. The method uses a fusion module to integrate the global and local features extracted by the two branches. The method is evaluated on five polyp segmentation datasets, and the results show that it achieves faster reasoning speed and higher accuracy than conventional CNN. The method was evaluated on CVC-ClinicDB, CVC-ColonDB, CVC-300, ETIS-LaribPolypDB and Kvasir-SEG.

A hybrid colorectal polyps segmentation network denoted as Fu-TransHNet was introduced by [47]. The model combines the global feature learning of Transformer and the local feature learning of CNN. Moreover, the model uses a novel fusion module to integrate the features from two branches and enhance the feature representation. The network is trained with a multi-view cooperation enhancement to obtain adaptive weights for each branch and the fusion module. The authors evaluate the model on five benchmark datasets: Kvasir-SEG, ETIS-LaribPolypDB, CVC-ColonDB, CVC-EndoScene and CVC-ClinicDB.

In [48], Chen et.al. introduced SegT, a novel polyp segmentation network capable of capturing both local and global information. The network utilizes a pyramid vision transformer (PVT) [49] as the encoder and a separated edge-guidance (SEG) module to refine the feature maps. The SEG module is composed of two blocks: a separator (SE) block that focuses on the foreground and background separately, and an edge-guidance (EG) block that fuses edge information into the features.  Moreover, a cascade fusion module (CFM) is presented to merge the refined multi-level features. SegT performance was evaluated on five public datasets: ETIS-LaribPolypDB, CVC-ClinicDB, CVC-ColonDB, CVC-300, and Kvasir-SEG.

\section{Proposed Method}
\subsection{Overview}
RetSeg follows the design of UNET and is fundamentally composed of three primary components: 1) An Encoder module responsible for feature extraction, 2) A Bottleneck segment incorporating the retention blocks, and 3) A Decoder component tasked with the generation of a mask from a provided image. RetSeg incorporates skip connections that establish direct pathways between corresponding layers in the Encoder and Decoder modules. These connections facilitate the transfer of detailed spatial information from the Encoder to the Decoder, thereby enhancing the precision of the output segmentation mask. In Figure 1, we present a high-level overview of RetSeg. Details on the specific design of each of the three modules are elaborated upon in the following subsections of this section.
\begin{figure}[htbp]
  \centering
  \includegraphics[width=0.8\textwidth]{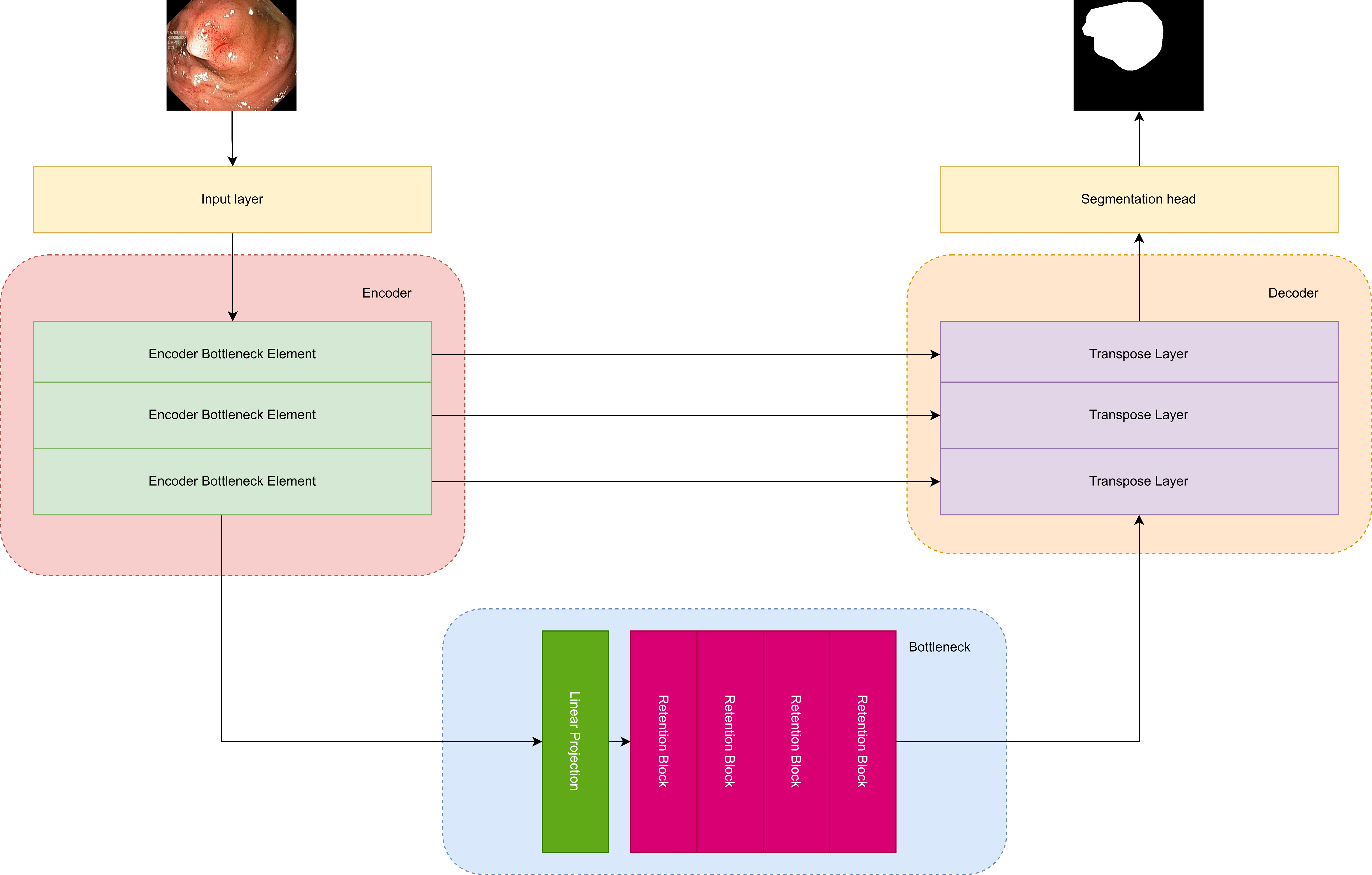}
  \caption{A high-level overview of the proposed RetSeg network}
  \label{fig:Fig1}
\end{figure}

\subsection{Data Collection}
We use five publicly accessible datasets to train and evaluate RetSeg. More precisely, the Kvasir-SEG and CVC-ClinicDB datasets are amalgamated to form the training and validation sets. Concurrently, the ETIS-LaribPolypDB, CVC-300, CVC-ColonDB, and BKAI-IGH NeoPolyp datasets are designated for testing procedures. The Kvasir-SEG dataset, a comprehensive collection of 1000 colonoscopy images, is accompanied by corresponding ground-truth masks. These images feature polyps of diverse shapes, colors, and sizes. Each sample has undergone annotation and verification by gastroenterologists.  The resolution of these samples spans from 332x487 to 1920x1072 pixels, with the entire set of 1000 samples being employed for training and validation purposes. The CVC-ClinicDB dataset is composed of 612 static polyp frames, extracted from a total of 31 colonoscopy sequences. All images maintain a uniform resolution of 384 × 288 pixels and have been extensively utilized for the testing and validation of various colorectal polyp segmentation methodologies.

Additionally, the CVC-300 dataset encompasses 300 images, each accompanied by a ground-truth mask and possessing a resolution of 574 x 500 pixels. These images are derived from a collection of 15 colonoscopy videos. The ETIS-LaribPolypDB dataset, on the other hand, consists of 196 samples with a resolution of 1225 × 996 pixels, captured under unfiltered conditions, with a subset of images exhibiting blur. The CVC-ColonDB dataset incorporates 300 static samples, a portion of which are captured under suboptimal conditions characterized by blur and challenging fields of vision. The samples in this dataset exhibit variation in size and shape.

Lastly, the BKAI-IGH NeoPolyp dataset comprises 1,200 samples along with their corresponding ground-truth masks. These samples are captured under challenging conditions and feature polyps of varying sizes. To summarize, we have compiled a robust dataset comprising 1612 samples. Of these, 20\% were allocated for validation while the remaining 80\% were utilized for training. In addition, we designated an extra set of 1996 samples specifically for testing. For consistency, we resized all our images to a constant dimensions of {224} $\times$ {224}. All samples used in this study are publicly available and intended solely for research purposes. A detailed distribution of the total sample count across each dataset, along with their respective purposes in this study, is provided in Table 1.

\begin{table}[htbp]
  \centering
  \caption{A breakdown of the datasets used in this study}
  \label{tab:Tab1}
  \begin{tabular}{|p{4cm}|p{2cm}|p{4cm}|} 
    \hline
    \textbf{Dataset} & \textbf{Size} & \textbf{Usage} \\
    \hline
    Kvasir-SEG & 1,000 & Training \\
    CVC-ClinicDB & 612 & Training \\
    CVC-300 & 300 & Testing \\
    CVC-ColonDB & 300 & Testing \\
   ETIS-LaribPolypDB & 196 & Testing \\
    BKAI-IGH NeoPolyp & 1,200 & Testing \\
    \hline
    \multicolumn{2}{|r|}{Total} & 3,608 \\
    \hline
  \end{tabular}
\end{table}

\subsection{Encoder Design}
Our encoder has been meticulously engineered to ensure efficient image processing and optimal feature extraction. It is composed of two fundamental modules: the Encoder Bottleneck Element (EBE) and the Processing Block (PB). The PB module is a key component of our encoder. It integrates depthwise and pointwise convolutional layers, which are activated using the Rectified Linear Unit (ReLU) activation function. To enhance the richness of the feature maps, we incorporate a residual connection to a feature aggregation layer. This connection creates a direct pathway from the input features to the downstream features, preserving important information throughout the network.  The EBE module is composed of multiple PB blocks arranged in a specific sequence. The features are initially processed by a PB block, then passed through a grouped convolution layer, and finally processed by another PB block. This sequence allows for complex transformations and interactions among the features. Furthermore, to capture a wider range of information, we fuse the features from the first and last PB blocks before they are passed forward. 

Throughout the encoder, we have opted for layer normalization in place of the commonly used batch normalization and have strategically incorporated dropout layers between each Encoder Bottleneck Element (EBE) module. Figure 2 provides an illustration of both EBE and PB blocks’ design.

\begin{figure}[htbp]
\centering
\includegraphics[width=1.0\textwidth]{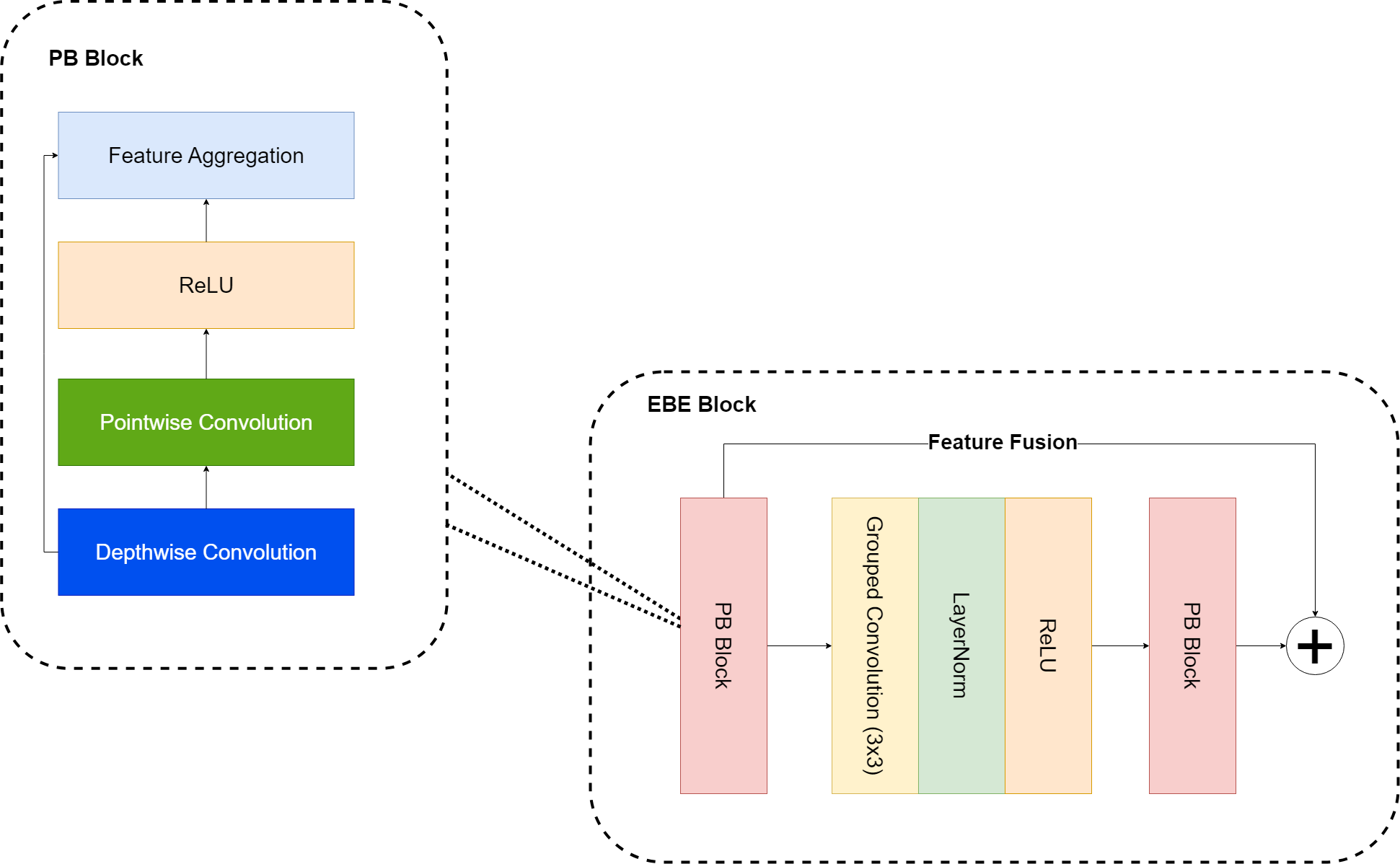}
\caption{The structure of a single encoder bottleneck element (EBE) block}
\label{Fig2}
\end{figure}
\subsection{Retention Bottleneck}
Our retention block comprises several components, some of which are closely similar to how vision transformers process images[49], [50]. At first, given an input image $x \in \mathbb{R}^{HWC}$ we attempt to transform it into a 1-D vector of shape $N \times D$ by dividing the image into $P \times P$ patches. Achieving this objective can be done using Equation 1:

\begin{equation}
N = \frac{HW}{P^2} \quad
\end{equation}

Each patch is then flattened and linearly transformed (tokenized) to create a sequence of tokens. Given a patch of shape $P \in \mathbb{R}^{PPC}$, we then can flatten it into vector $v \in \mathbb{R}^{(P^2c)}$. Using an embedding matrix, we then linearly transform the patches using a learnable embedding matrix $E \in \mathbb{R}^{P^2C\times D}$. Using Equation 2, we can apply this transformation:

\begin{equation}
x = E \times v
\end{equation}

This process results in creating a sequence of tokens \(X=[x_1,x_2,\ldots,x_N]\) where \(x_i \in \mathbb{R}^D\). The sequence of tokens is subsequently processed by our multi-head retention block. In order to apply the retention mechanism to our tokens, we adopt the RMT design as referenced in [22] where the retention is implemented as bidirectional rather than unidirectional. Given a recurrent sequence of tokens, represented as:

\begin{equation}
o_n = \sum_{m=1}^{N} \gamma^{|n-m|} (Q_n e^{in\theta})(K_m e^{im\theta})^{\dagger} v_m \quad
\end{equation}

Where \(N\) is the number of tokens. We can then apply retention using Equation 4:

\begin{equation}
R(X) = Q \cdot K^T, \quad R(X) = (R(X) \odot D^2d) V \quad
\end{equation}

Where \(Q\), \(K\), \(V\) and \(D\) are defined as:

\begin{equation}
Q = (XW_Q) \odot \Theta, \quad K = (XW_K) \odot \bar{\Theta}, \quad V = XW_V
\end{equation}

\[
\Theta_n = e^{in\theta}, \quad \mathbb{D}_{nm}^{2d} = \gamma^{|x_n-x_m|+|y_n-y_m|} \quad
\]
In Equation 5, \(\bar{\Theta}\) represents the complex conjugate of \(\Theta\), and \(\mathbb{D}_{nm}^{2d} \in \mathbb{R}^D\) is a decay mask which represents the relative distance between token pairs. In contrast to the original and RMT implementations, we do not employ group normalization or softmax, as these were found to diminish our model’s performance. Instead, we opt for layer normalization applied to the output of the retention block. In Figure 3, we present the overall structure of our retention block.

\begin{figure}[htbp]
  \centering
  \includegraphics[width=0.5\textwidth]{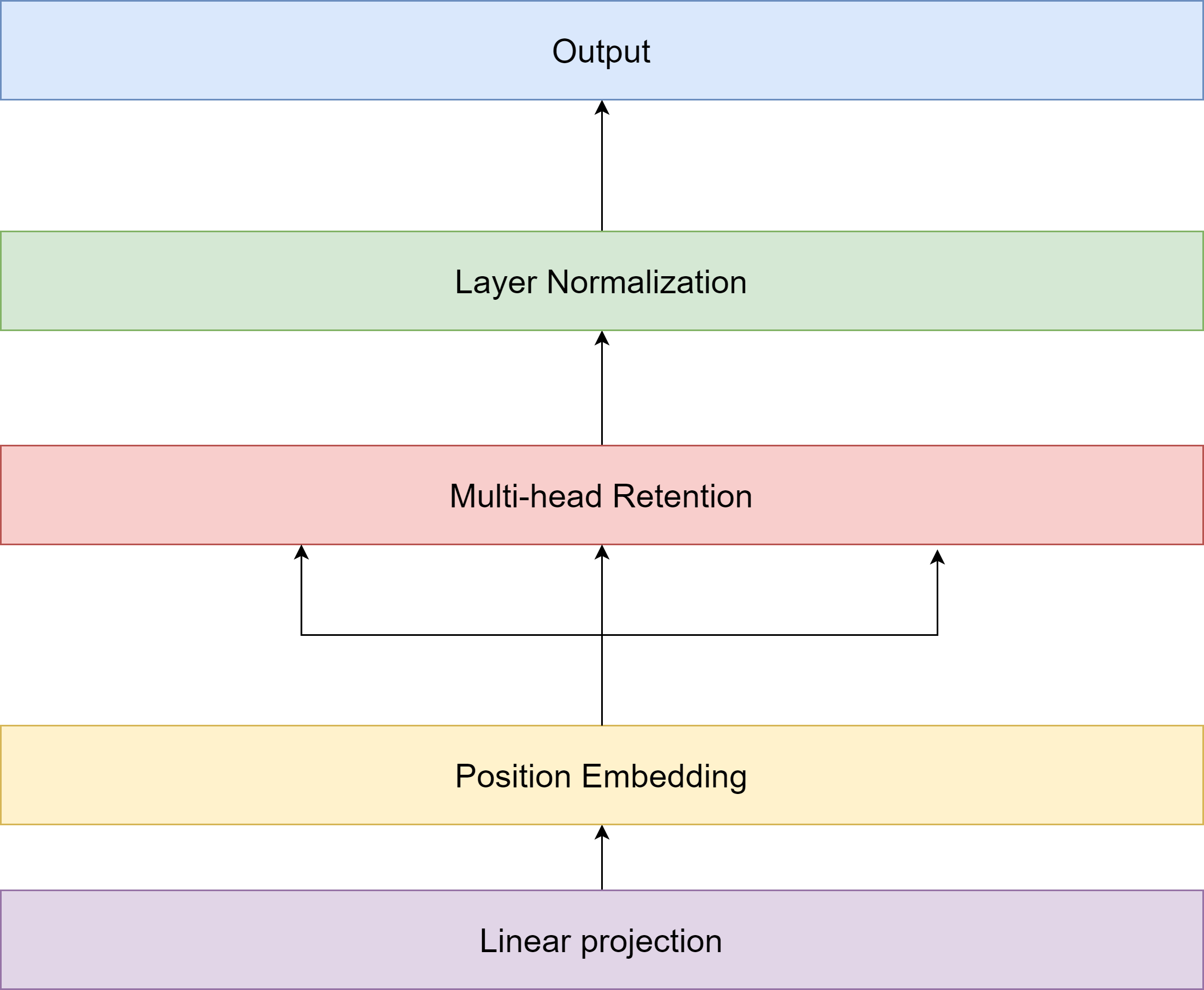}
  \caption{The design of a single retention block. Our retention bottleneck comprises several retention blocks}
  \label{fig:Fig2}
\end{figure}

\subsection{Decoder}
Our decoder comprises several custom modules, termed referred to as decoder bottlenecks, each of which includes an upsampling operation and a sequence of convolutional layers and activation functions. The upsampling operation doubles the spatial dimensions of the input feature map, utilizing bilinear interpolation. Following this, a series of convolutional layers are applied. These layers are activated using the ReLU activation functions, which are applied in-place to conserve memory. The decoder bottlenecks are arranged sequentially, each serving to reduce the number of channels while simultaneously increasing the spatial dimensions of the feature maps.  An additional convolutional layer is incorporated at the end of the Decoder module, followed by a sigmoid function. This function maps the output to two distinct classes: polyps and the background. This is particularly useful in binary segmentation tasks where each pixel is classified into one of these two categories.

\subsection{Loss functions}
In our approach to optimize RetSeg, we employ a blend of loss functions, specifically binary cross entropy, Focal Loss, and Reconstruction loss. Weighted coefficients, denoted as \(\alpha\) and  \(\beta\) , are incorporated into the binary cross entropy and focal losses. By fine-tuning these coefficients, we can adjust the relative significance of the binary cross-entropy loss and focal loss. This method of balancing the two loss functions has proven effective in enhancing the performance of our model on the task of polyp segmentation. Our total loss can be represented in Equation 6:

\begin{equation}
L_{{total}} = \alpha L_{{BCE}} + \beta L_{{focal}} + L_{{recon}} \quad 
\end{equation}

Each loss function employed in the training process of RetSeg serves a distinct purpose. Binary cross-entropy, for instance, quantifies the divergence between the predicted probabilities and the actual class labels in binary classification tasks. On the other hand, Focal Loss addresses the issue of class imbalance by assigning lesser weights to examples that are easier to classify. Lastly, L1 loss, which calculates the absolute differences between the predicted and true values, aids in enhancing the precision of the predictions.

\section{Experimental Results}
\subsection{Overview}
To assess the performance of RetSeg under various conditions, we executed a set of experiments on four unique datasets. During both the training and inference phases, we kept track of several metrics, including Intersection over Union (IoU), Dice coefficient, precision, recall, F1 score, and mean squared error. The model was trained and tested on a standalone machine outfitted with a single RTX5000 GPU and running the Linux operating system.   We showcase a comparative analysis between RetSeg and prevalent segmentation models, in addition to evaluating the performance of RetSeg against recently proposed methods for polyp segmentation. It’s important to note that these experimental results are preliminary. It is crucial to note that this study is still in its early stages. The experimental results obtained so far are preliminary. As we progress, further investigations will be carried out to comprehend the correlation between different components of RetSeg and the overall performance of the model. These future studies will provide more insights and potentially lead to improvements in the model’s performance.

\subsection{Performance Analysis}
In this subsection, we assess RetSeg’s performance across four distinct datasets. A detailed analysis is conducted using various metrics including Intersection over Union (IoU), Dice Coefficient, Precision, Recall, F-score and Mean Squared Error (MSE). In addition, we examine the frames-per-second of our model on each dataset. The results of this analysis are presented in Table 2. Moreover, we present the training and validation loss and accuracies in Figure 4. This involves showcasing a number of accurate and inaccurate predictions made by RetSeg. The insights from this analysis are visually represented in Figures 5 and 6.

\begin{table}[htbp]
\centering
\caption{RetSeg Performance on Different Datasets}
\label{tab:Tab2}
\begin{tabular}{lccccccc}
\toprule
\textbf{Dataset} & \textbf{IoU} & \textbf{Dice} & \textbf{Precision} & \textbf{Recall} & \textbf{F1} & \textbf{MSE} & \textbf{FPS}\\
\midrule
ETIS-LaribPolypDB & 0.964 & 0.981 & 0.986 & 0.977 & 0.982 & 0.015 & 55.33\\
CVC-300 & 0.951 & 0.975 & 0.995 & 0.955 & 0.975 & 0.020 & 90.04\\
CVC-ColonDB & 0.923 & 0.960 & 0.989 & 0.932 & 0.960 & 0.035 & 93.20\\
BKAI-IGH NeoPolyp & 0.958 & 0.978 & 0.983 & 0.973 & 0.978 & 0.017 & 89.95\\

\bottomrule
\end{tabular}
\end{table}

\begin{figure}[htbp]
\centering
\includegraphics[width=0.60\textwidth]{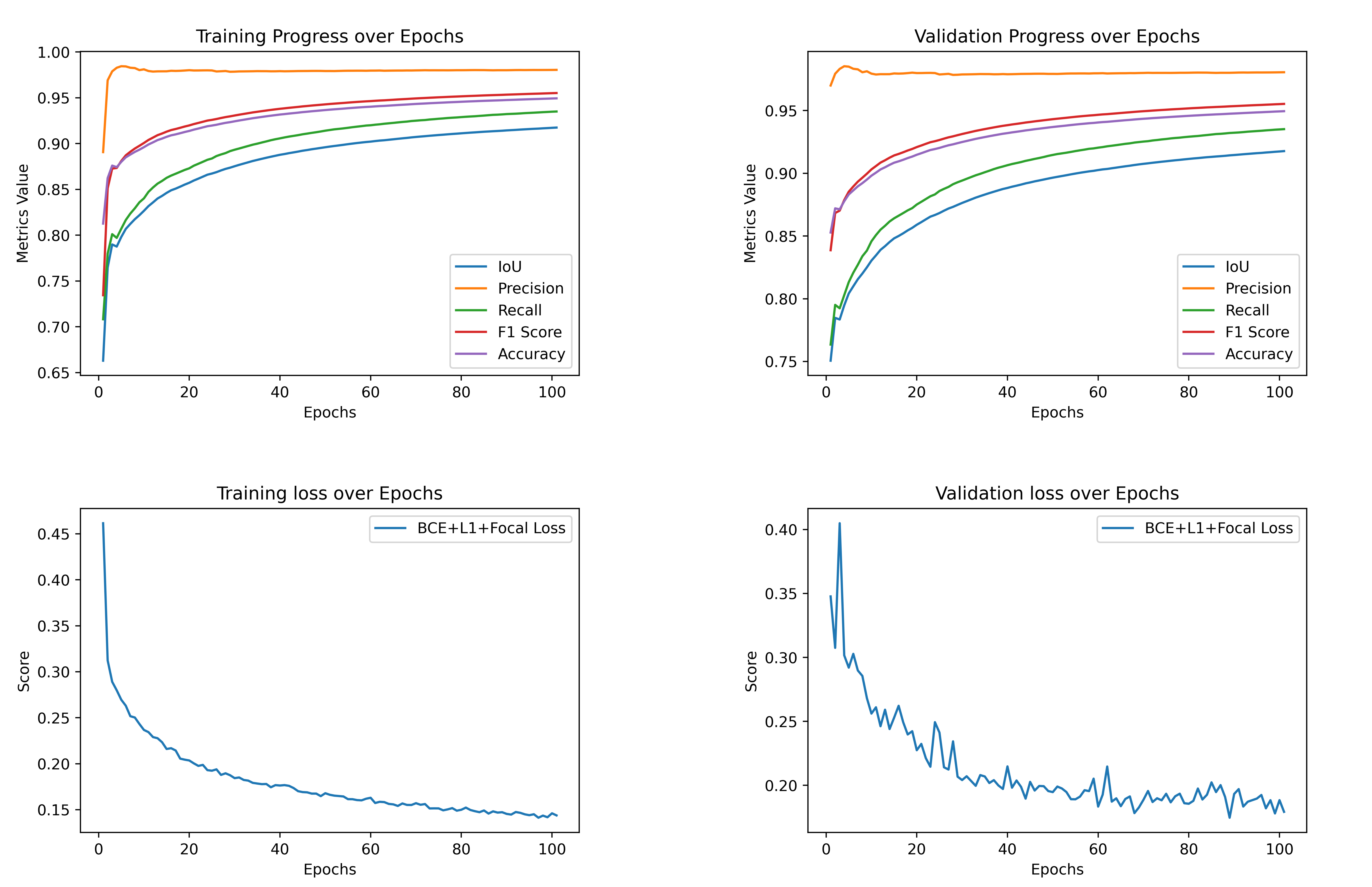}
\caption{RetSeg's progression of training and validation loss, along with accuracies, over the course of 100 epochs}
\label{graphs}
\end{figure}

\begin{figure}[htbp]
\centering
\includegraphics[width=1\textwidth]{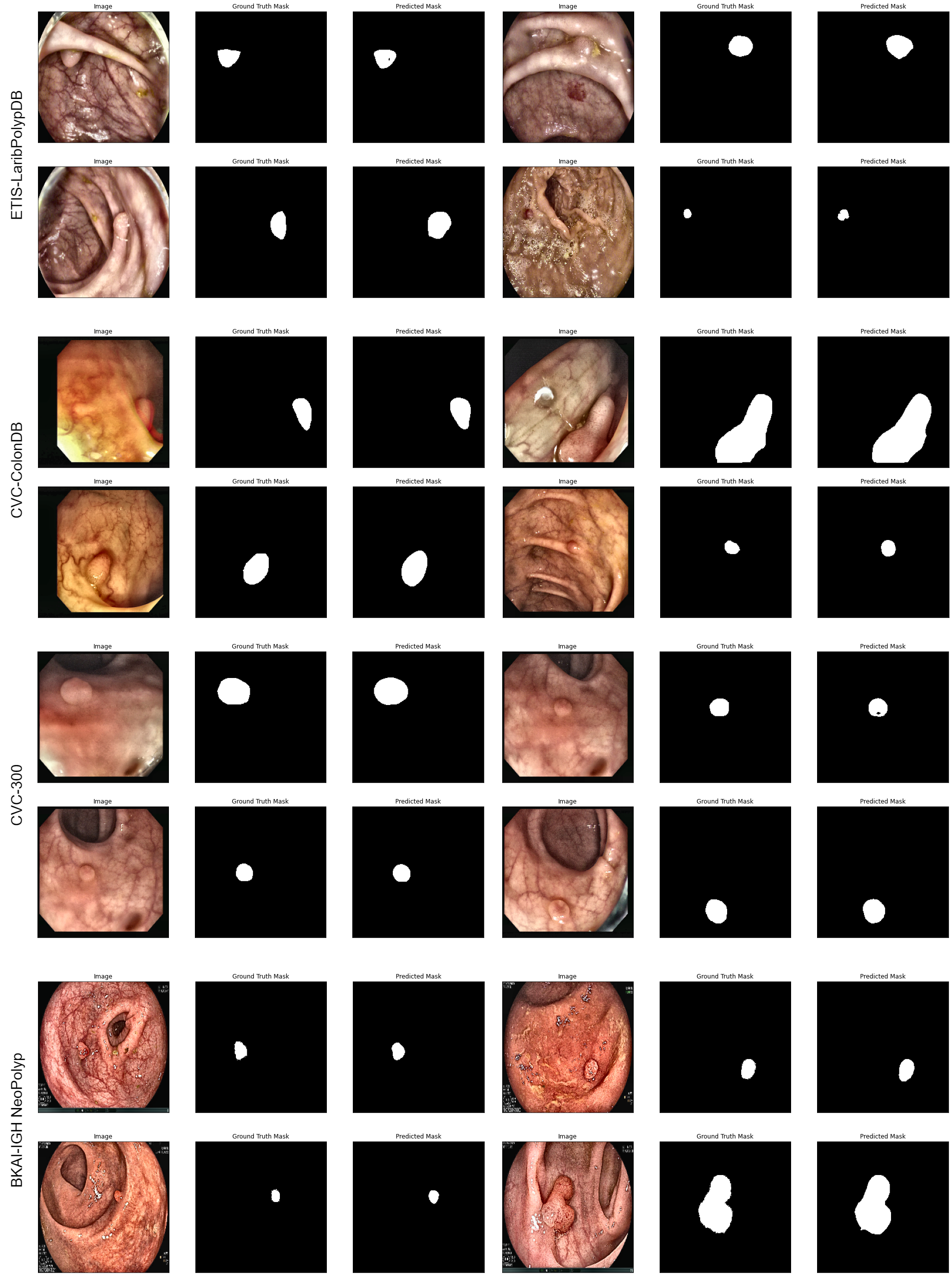}
\caption{Several examples of correct predictions}
\label{correct}
\end{figure}

\begin{figure}[htbp]
\centering
\includegraphics[width=1\textwidth]{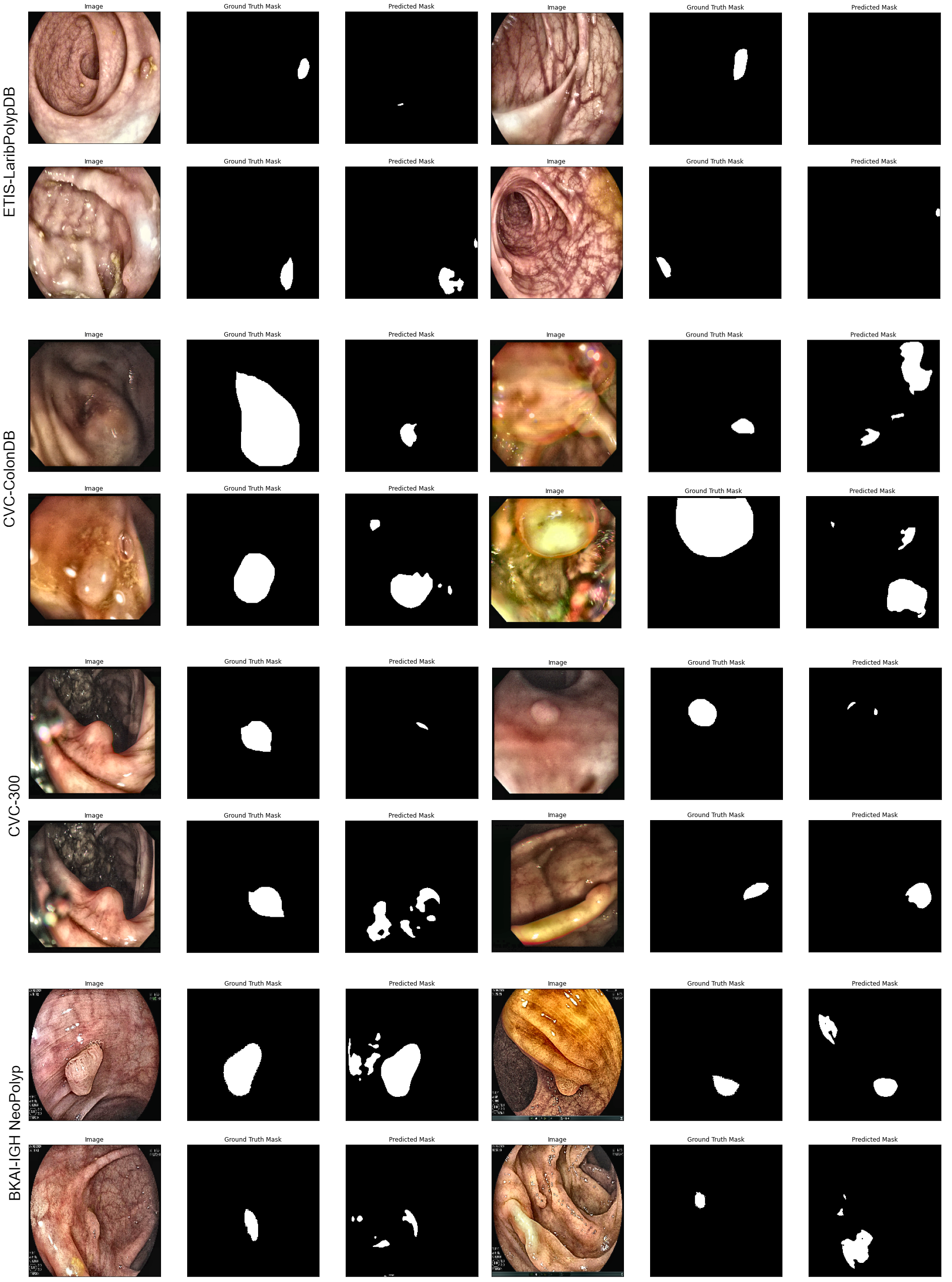}
\caption{Several examples of failed predictions}
\label{failed}
\end{figure}

\subsection{Comparative Analysis}
We evaluate the performance of RetSeg in comparison to several prevalent segmentation models, including UNET, UNET++ and DeepLabV3. Furthermore, we compare RetSeg’s performance with that of recently introduced polyp segmentation methods to gain a deeper understanding of the performance differences. In Table 3, we showcase a comparative analysis of RetSeg’s performance with the abovementioned segmentation models. Each model is pre-trained and evaluated on each of the four testing datasets.

\begin{table}[htbp]
\centering
\caption{Model Performance Metrics. The best metrics are in bold}
\label{tab:model_metrics}
\renewcommand{\arraystretch}{1.5} 

\begin{tabularx}{\textwidth}{|X|X|c|c|c|c|c|c|c|}
\hline
\textbf{Model} & \textbf{Dataset} & \textbf{IoU} & \textbf{Dice} & \textbf{Precision} & \textbf{Recall} & \textbf{F1} & \textbf{MSE} \\
\hline
\multirow{4}{*}{RetSeg} & ETIS-LaribPolypDB & {0.964} & {0.981} & 0.986 & 0.977 & {0.982} & {0.015}\\
 & CVC-ColonDB & {0.923} & {0.960} & 0.989 & 0.932 & {0.960} & {0.035}\\
 & CVC-300 & {0.951} & {0.975} & {0.995} & 0.955 & {0.975} & {0.020} \\
 & BKAI-IGH NeoPolyp & {0.958} & {0.978} & 0.983 & 0.973 & {0.978} & {0.017} \\
\hline
\multirow{4}{*}{UNET} & ETIS-LaribPolypDB & 0.698 & 0.816 & 0.858 & 0.724 & 0.881 & 0.010 \\
 & CVC-ColonDB & 0.826 & 0.909 & 0.921 & 0.891 & 0.903 & 0.034 \\
 & CVC-300 & 0.720 & 0.699 & 0.982 & {0.983} & {0.983} & {0.015} \\
 & BKAI-IGH NeoPolyp & 0.869 & 0.851 & 0.913 & 0.921 & 0.905 & 0.039 \\
\hline
\multirow{4}{*}{UNET++} & ETIS-LaribPolypDB & 0.875 & 0.861 & 0.979 & 0.919 & 0.973 & 0.014 \\
 & CVC-ColonDB & 0.815 & 0.796 & 0.913 & 0.907 & 0.975 & 0.045 \\
 & CVC-300 & {0.873} & {0.841} & {0.987} & {0.991} & {0.989} & {0.009} \\
 & BKAI-IGH NeoPolyp & 0.887 & 0.869 & 0.919 & 0.951 & 0.969 & 0.026 \\
\hline
\multirow{4}{*}{DeepLabV3} & ETIS-LaribPolypDB & 0.870 & 0.844 & 0.979 & {0.963} & {0.970} & 0.054 \\
 & CVC-ColonDB & 0.825 & 0.814 & 0.915 & 0.889 & 0.881 & 0.058 \\
 & CVC-300 & 0.811 & 0.792 & 0.989 & {0.988} & {0.989} & {0.008} \\
 & BKAI-IGH NeoPolyp & {0.912} & {0.900} & {0.993} & 0.953 & {0.972} & {0.023} \\
\hline
\end{tabularx}
\end{table}

We have conducted a comparative analysis between RetSeg and several other recently proposed polyp segmentation methods. To maintain fairness in this comparison, we have utilized the same datasets that were used in these prior studies to conduct the analysis. The results of this analysis are elaborated in Tables 4 and 5.

\begin{table}[htbp]
\centering
\caption{Performance Comparison of RetSeg with Recently Proposed Methods}
\label{tab:model_metrics_comparison}
\renewcommand{\arraystretch}{1.5} 
\begin{tabularx}{\textwidth}{|X|X|c|c|c|c|}
\hline
\textbf{Model} &  \textbf{Dataset} &  \textbf{IoU} &  \textbf{Dice} \\
\hline
RetSeg & \multirow{9}{*}{ETIS-LaribPolypDB} & 0.964 & 0.981 \\
\cline{1-1}\cline{3-4}
ESFPNet [51] & & {0.752} & {0.827} \\
\cline{1-1}\cline{3-4}
Cascading Attn. [52] & & 0.725 & 0.800 \\
\cline{1-1}\cline{3-4}
DUCKNet [53] & & 0.905 & 0.950 \\
\cline{1-1}\cline{3-4}
ESSL-Polyp [54] & & 0.776 & 0.705 \\
\cline{1-1}\cline{3-4}
CAFE-NET[55] & & 0.738 & 0.822 \\
\cline{1-1}\cline{3-4}
Patch Network[56] & & 0.940 & 0.964 \\
\cline{1-1}\cline{3-4}
Coarse-to-fine [57] & & 0.616 & 0.548 \\
\cline{1-1}\cline{3-4}
DuAT [40] & & 0.746 & 0.822 \\
\hline
\end{tabularx}
\end{table}

\begin{table}[H] 
\centering
\caption{Performance Comparison of RetSeg with Recently Proposed Methods (continued)}
\label{tab:model_metrics2} 
\renewcommand{\arraystretch}{1.5} 
\begin{tabularx}{\textwidth}{|X|X|c|c|c|c|}
\hline
\textbf{Model} &  \textbf{Dataset} &  \textbf{IoU} &  \textbf{Dice} \\
\hline
RetSeg & \multirow{8}{*}{CVC-ColonDB} & 0.923 & 0.960 \\
\cline{1-1}\cline{3-4}
ResUNet++ [58] & & 0.708 & 0.568 \\
\cline{1-1}\cline{3-4}
PolypPVT [59] & & 0.808 & 0.727 \\
\cline{1-1}\cline{3-4}
SSFormer [60] & & 0.721 & 0.802 \\
\cline{1-1}\cline{3-4}
Multi Parallel UNET [61] & & 0.653 & 0.770 \\
\cline{1-1}\cline{3-4}
Poly-SAM [62] & & 0.891 & 0.894 \\
\cline{1-1}\cline{3-4}
SR-AttNet[63] & & 0.539 & 0.665 \\
\cline{1-1}\cline{3-4}
MetaPolyp [64] & & 0.867 & 0.790 \\
\hline
RetSeg & \multirow{7}{*}{BKAI-IGH NeoPolyp} & 0.958 & 0.978 \\
\cline{1-1}\cline{3-4}
Focal-UNet[65] & & - & {0.802} \\
\cline{1-1}\cline{3-4}
TransResU-Net [25] & & 0.856 & 0.915 \\
\cline{1-1}\cline{3-4}
RaBiT [66] & & 0.886 & 0.940 \\
\cline{1-1}\cline{3-4}
DilatedSegNet [67] & & 0.831 & 0.895 \\
\cline{1-1}\cline{3-4}
TGANet [68] & & 0.840 & 0.902 \\
\cline{1-1}\cline{3-4}
NeoUnet [69] & & 0.837 & 0.911 \\
\hline
RetSeg & \multirow{4}{*}{CVC-300} & 0.951 & 0.975 \\
\cline{1-1}\cline{3-4}
HarDNet + UNET [70] & & 0.857 & {0.923} \\
\cline{1-1}\cline{3-4}
Mobile-PolypNet[71] & & 0.864 & 0.901 \\
\cline{1-1}\cline{3-4}
Pseudo labeling [72] & & 0.836 & 0.904 \\
\hline
\end{tabularx}
\end{table}

\section{Discussion}
\subsection{RetSeg Design}
The design of our network was carefully designed a series of experiments, aimed at optimizing the extraction of valuable features from the input data. We adopted a U-NET shaped architecture, a well-regarded framework known for its efficacy in image segmentation tasks. In this architecture, we strategically established skip connections between the encoder and decoder layers, facilitating the seamless flow of information and ensuring the creation of an enriched feature map.
Furthermore, to enhance the model's feature extraction capabilities, we integrated both depthwise and pointwise convolutions within the processing block (PB). This combination of convolutional operations in the PB played a critical role in capturing intricate patterns and structures in the input data. The depthwise convolutions efficiently captured spatial information, while the subsequent pointwise convolutions condensed this information, striking a balance between computational efficiency and feature richness.  

Moreover, we implemented a features aggregation mechanism to consolidate and integrate information across different layers of the network. This aggregation process augmented the model's understanding of complex features by aggregating high-level abstractions from various stages of the network. The resulting feature map became a synthesis of comprehensive representations, contributing to improved segmentation performance. The model’s encoder bottleneck element block (EBE) has also been carefully developed to capture an extensive array of valuable features. This is achieved through the composition of multiple processing block (PB) blocks within the EBE, each designed to maximize feature extraction.  The inclusion of grouped convolutions between the PB blocks further enhances the receptive field and encourages the model to differentiate complex patterns within the feature maps. 

In addition to grouped convolutions, we introduced feature fusion mechanisms within the EBE. These mechanisms play a pivotal role in consolidating the features obtained from different PB blocks, promoting a holistic representation of the input information. By fusing these diverse features, the EBE ensures that the subsequent layers of the network receive a rich and varied set of features. In relation to the retention blocks, our experimental findings strongly support their integration within the bottleneck section of RetSeg as the most effective approach for our specific case. 

We hypothesize that incorporating the retention mechanism within this bottleneck has significantly aided the model in giving priority to discriminative and finely-refined features. Moreover, situating the retention mechanism at this strategic location within RetSeg facilitated the modeling of global context at a higher level of abstraction. 
\vspace{-1em} 
\subsection{Retention Mechanism}
While RetSeg comprises several distinguishing characteristics compared to existing methods, its primary differentiation stems from the introduction of the retention mechanism. This approach has showcased notable effectiveness, although further research is warranted for comprehensive validation. The retention mechanism used in RetSeg focuses on utilizing a decay factor to model a specific 2-dimensional distance. This decay factor is calculated based on the properties of the input data and helps in capturing meaningful spatial relationships within the image. The decay factor is then incorporated through a multiplication operation with a weight matrix, enhancing the model's ability to emphasize critical features based on spatial importance. Moreover, an important aspect of this retention mechanism is its approach to token processing. Unlike attention-based methods that process one token at a time, the retention mechanism processes tokens in sequences, allowing for efficient and simultaneous processing of multiple tokens as previously reported in [22].

This not only conserves computational resources but also ensures optimized performance. By handling multiple tokens simultaneously, the algorithm can capture more complex patterns and dependencies within the data, ultimately leading to more accurate segmentation results. This token processing strategy demonstrates the model's capacity to handle and extract intricate information from the input, further underlining the effectiveness of the retention mechanism in the context of polyp segmentation We speculate that incorporating chunkwise and recurrent retention mechanisms could significantly enhance the performance of RetSeg. However, conducting meticulous experiments is essential to fully leverage the potential strength of the retention mechanisms.
\vspace{-1em} 
\subsection{Result Interpretation and Limitations}
Throughout our experiments, we have identified several interesting patterns. Primarily, upon careful examination of the results presented in Table 3, it becomes apparent that RetSeg exhibits generally higher Intersection over Union (IoU), dice coefficient, and F1 scores in comparison to other existing models. However, it is noteworthy that certain metrics, particularly precision and recall, tend to register comparatively lower values when contrasted with those of other segmentation models.  

We theorize that such phenomena occur as a result of RetSeg’s bias towards achieving a higher true positive rate, consequently prioritizing accurate identification of polyp regions even at the expense of potentially including false positives.  This inclination towards over-identification may inflate the IoU, dice coefficient, and F1 scores, as the model is successful in capturing a significant portion of the actual polyp regions. However, this strategy could compromise precision, as a larger area identified as a polyp might encompass non-polyp regions. Similarly, recall may suffer because some actual polyp areas might be misclassified or omitted due to the model's broad segmentation approach.

Moreover, upon careful examination of both the correct and incorrect predictions showcased in Figure 5 and 6, we observe that RetSeg demonstrates a noteworthy ability in handling a wide array of challenging shapes and sizes of polyps. In Figure 5, the correct predictions demonstrate the model's ability to accurately segment polyps even when their shapes vary considerably, highlighting its versatility in capturing diverse anatomical features.

However, a closer analysis of the failed predictions in Figure 6 unveils a limitation of RetSeg in identifying polyps under extreme conditions. Particularly, RetSeg struggles when faced with challenging fields of vision, such as distorted perspectives or unusual camera angles. Additionally, RetSeg fails to produce an accurate mask when presented with blurred images, which can be a common occurrence in real-world endoscopic scenarios. In such cases, the blurriness can distort the boundary of the polyp, making accurate segmentation a formidable task for the model.

Furthermore, RetSeg encounters difficulties processing images with additional obstructions, such as white light reflections. These obstructions can interfere with the model's ability to accurately identify and delineate the true boundaries of the polyps. Consequently, the presence of these adverse conditions poses a substantial challenge for RetSeg, impacting its overall segmentation performance.

As previously stated, our ongoing work necessitates further investigation to refine and potentially enhance RetSeg’s performance. A significant constraint of this study is the scarcity of research on the application of chunkwise and recurrent retention within RetSeg. Moreover, it’s crucial to undertake additional studies with real-world samples for an exhaustive evaluation of RetSeg’s performance. This becomes particularly significant when considering devices with diverse specifications, given our ultimate goal is to deploy this model onto actual colonoscopes.

While Figure 5 showcases correct predictions that highlight the model’s ability to accurately segment polyps despite considerable variations in their shapes, thereby demonstrating its versatility in capturing a wide range of anatomical features, a detailed analysis of the failed predictions in Figure 6 reveals a limitation of RetSeg in identifying polyps under extreme conditions.

Specifically, RetSeg faces challenges when dealing with difficult fields of vision, such as distorted perspectives or unusual camera angles. Furthermore, RetSeg struggles to generate an accurate mask when confronted with blurred images, a situation that can frequently occur in real-world endoscopic scenarios. In these instances, the blurriness can distort the polyp’s boundary, making accurate segmentation a daunting task for the model. Additionally, RetSeg has difficulties processing images with obstructions like white light reflections. These obstructions can hinder the model’s ability to accurately identify and outline the true boundaries of the polyps. As a result, these adverse conditions present a significant challenge for RetSeg, affecting its overall segmentation performance.

\section{Conclusion}
In this study, we present RetSeg, a retention-based segmentation network designed specifically for the detection of colorectal polyps. Our approach is inspired by the recent success of Retentive Networks, which have showcased superior performance compared to transformers in natural language processing (NLP) tasks. We introduce a  multi-head retention block and integrate it into the bottleneck portion of the network. So far in this study we have only utilized the parallel representation paradigm for our multi-head retention blocks; however, we plan to conduct further experiments to explore the potential of using recurrent and chunkwise representations paradigms. To optimize our network , we utilize a combination of loss functions, including binary cross-entropy, Dice loss, focal loss, and L1 loss. Extensive testing of RetSeg on various datasets demonstrates its promising performance in terms of both efficiency and accuracy when compared to existing segmentation networks. Notably, RetSeg upholds the fundamental properties of retentive networks, such as training parallelism and lower resources consumption, reinforcing its viability and effectiveness in medical image segmentation tasks. Although the findings are promising, further studies are required to enhance the integration of retention into segmentation tasks.

\bibliographystyle{plain}

\begin{thebibliography}{72}

\bibitem{author1}
A. Vaswani et al., “Attention is all you need,” Adv Neural Inf Process Syst, vol. 2017-Decem, no. Nips, pp. 5999–6009, 2017.

\bibitem{author2}
D. Khurana, A. Koli, K. Khatter, and S. Singh, “Natural language processing: state of the art, current trends and challenges,” Multimed Tools Appl, vol. 82, no. 3, pp. 3713–3744, 2023, doi: 10.1007/s11042-022-13428-4.

\bibitem{author3}
G. Yenduri et al., “Generative Pre-trained Transformer: A Comprehensive Review on Enabling Technologies, Potential Applications, Emerging Challenges, and Future Directions,” pp. 1–40, 2023, [Online]. Available: http://arxiv.org/abs/2305.10435

\bibitem{author4}
T. Lin, Y. Wang, X. Liu, and X. Qiu, “A survey of transformers,” AI Open, vol. 3, no. September, pp. 111–132, 2022, doi: 10.1016/j.aiopen.2022.10.001.

\bibitem{author5}
A. Dosovitskiy et al., “an Image Is Worth 16X16 Words: Transformers for Image Recognition At Scale,” ICLR 2021 - 9th International Conference on Learning Representations, 2021.

\bibitem{author6}
S. Khan, M. Naseer, M. Hayat, S. W. Zamir, F. S. Khan, and M. Shah, “Transformers in Vision: A Survey,” ACM Comput Surv, vol. 54, no. 10, pp. 1–30, 2022, doi: 10.1145/3505244.

\bibitem{author7}
K. Han et al., “A Survey on Vision Transformer,” IEEE Trans Pattern Anal Mach Intell, vol. 45, no. 1, pp. 87–110, 2023, doi: 10.1109/TPAMI.2022.3152247.

\bibitem{author8}
S. Cuenat and R. Couturier, “Convolutional Neural Network (CNN) vs Vision Transformer (ViT) for Digital Holography,” 2022 2nd International Conference on Computer, Control and Robotics, ICCCR 2022, pp. 235–240, 2022, doi: 10.1109/ICCCR54399.2022.9790134.

\bibitem{author9}
K. Al-hammuri, F. Gebali, A. Kanan, and I. T. Chelvan, “Vision transformer architecture and applications in digital health: a tutorial and survey,” Vis Comput Ind Biomed Art, vol. 6, no. 1, 2023, doi: 10.1186/s42492-023-00140-9.

\bibitem{author10}
E. U. Henry, O. Emebob, and C. A. Omonhinmin, “Vision Transformers in Medical Imaging: A Review,” 2022, [Online]. Available: http://arxiv.org/abs/2211.10043

\bibitem{author11}
F. Shamshad et al., “Transformers in medical imaging: A survey,” Med Image Anal, vol. 88, no. March, p. 102802, 2023, doi: 10.1016/j.media.2023.102802.

\bibitem{author12}
A. He, K. Wang, T. Li, C. Du, S. Xia, and H. Fu, “H2Former: An Efficient Hierarchical Hybrid Transformer for Medical Image Segmentation,” IEEE Trans Med Imaging, vol. 42, no. 9, pp. 2763–2775, 2023, doi: 10.1109/TMI.2023.3264513.

\bibitem{author13}
Q. Liu, C. Kaul, J. Wang, C. Anagnostopoulos, R. Murray-Smith, and F. Deligianni, “Optimizing Vision Transformers for Medical Image Segmentation,” vol. 104690, pp. 1–5, 2023, doi: 10.1109/icassp49357.2023.10096379.

\bibitem{author14}
A. Chernyavskiy, D. Ilvovsky, and P. Nakov, “Transformers: ‘The End of History’ for NLP?,” 2021, [Online]. Available: http://arxiv.org/abs/2105.00813

\bibitem{author15}
N. Patwardhan, S. Marrone, and C. Sansone, “Transformers in the Real World: A Survey on NLP Applications,” Information (Switzerland), vol. 14, no. 4, 2023, doi: 10.3390/info14040242.

\bibitem{author16}
L. H. Mormille, C. Broni-Bediako, and M. Atsumi, “Regularizing self-attention on vision transformers with 2D spatial distance loss,” Artif Life Robot, vol. 27, no. 3, pp. 586–593, 2022, doi: 10.1007/s10015-022-00774-7.

\bibitem{author17}
K. Kim et al., “Rethinking the self-attention in vision transformers,” IEEE Computer Society Conference on Computer Vision and Pattern Recognition Workshops, pp. 3065–3069, 2021, doi: 10.1109/CVPRW53098.2021.00342.

\bibitem{author18}
J. Yang et al., “Focal Attention for Long-Range Interactions in Vision Transformers,” Adv Neural Inf Process Syst, vol. 36, pp. 30008–30022, 2021.

\bibitem{author19}
P. Mehrani and J. K. Tsotsos, “Self-attention in vision transformers performs perceptual grouping, not attention,” Front Comput Sci, vol. 5, pp. 1–30, 2023, doi: 10.3389/fcomp.2023.1178450.

\bibitem{author20}
L. Wu et al., “Demystify Self-Attention in Vision Transformers from a Semantic Perspective: Analysis and Application,” 2022, [Online]. Available: http://arxiv.org/abs/2211.08543

\bibitem{author21}
Y. Sun et al., “Retentive Network: A Successor to Transformer for Large Language Models,” pp. 1–14, 2023, [Online]. Available: http://arxiv.org/abs/2307.08621

\bibitem{author22}
Q. Fan, H. Huang, M. Chen, H. Liu, and R. He, “RMT: Retentive Networks Meet Vision Transformers,” 2023, [Online]. Available: http://arxiv.org/abs/2309.11523

\bibitem{author23}
O. Ronneberger, P. Fischer, and T. Brox, “U-Net: Convolutional Networks for Biomedical Image Segmentation”, Accessed: Dec. 21, 2022. [Online]. Available: http://lmb.informatik.uni-freiburg.de/

\bibitem{author24}
J. Chen et al., “TransUNet: Transformers Make Strong Encoders for Medical Image Segmentation,” pp. 1–13, 2021, [Online]. Available: http://arxiv.org/abs/2102.04306

\bibitem{author25}
N. K. Tomar, A. Shergill, B. Rieders, U. Bagci, and D. Jha, “TransResU-Net: Transformer based ResU-Net for Real-Time Colonoscopy Polyp Segmentation,” pp. 1–4, 2022, [Online]. Available: http://arxiv.org/abs/2206.08985

\bibitem{author26}
D. Jha, P. H. Smedsrud, and M. A. Riegler, “Kvasir-SEG: A Segmented Polyp Dataset,” vol. 2, pp. 451–462, doi: 10.1007/978-3-030-37734-2.

\bibitem{author27}
J. Bernal, J. Sánchez, and F. Vilariño, “Towards automatic polyp detection with a polyp appearance model,” in Pattern Recognition, Sep. 2012, pp. 3166–3182. doi: 10.1016/j.patcog.2012.03.002.

\bibitem{author28}
P. M. Colucci, S. H. Yale, and C. J. Rall, “Colorectal polyps.,” Clin Med Res, vol. 1, no. 3, pp. 261–262, 2003, doi: 10.3121/cmr.1.3.261.

\bibitem{author29}
Y. Hao, Y. Wang, M. Qi, X. He, Y. Zhu, and J. Hong, “Risk factors for recurrent colorectal polyps,” Gut Liver, vol. 14, no. 4, pp. 399–411, 2020, doi: 10.5009/gnl19097.

\bibitem{author30}
M. Øines, L. M. Helsingen, M. Bretthauer, and L. Emilsson, “Epidemiology and risk factors of colorectal polyps,” Best Pract Res Clin Gastroenterol, vol. 31, no. 4, pp. 419–424, 2017, doi: 10.1016/j.bpg.2017.06.004.

\bibitem{author31}
T. K. L. Lui and W. K. Leung, “Is artificial intelligence the final answer to missed polyps in colonoscopy?,” World J Gastroenterol, vol. 26, no. 35, pp. 5248–5255, 2020, doi: 10.3748/WJG.V26.I35.5248.

\bibitem{author32}
W. Latos et al., “Colonoscopy: Preparation and Potential Complications,” Diagnostics, vol. 12, no. 3, pp. 1–10, 2022, doi: 10.3390/diagnostics12030747.

\bibitem{author33}
N. H. Kim et al., “Miss rate of colorectal neoplastic polyps and risk factors for missed polyps in consecutive colonoscopies,” Intest Res, vol. 15, no. 3, pp. 411–418, 2017, doi: 10.5217/ir.2017.15.3.411.

\bibitem{author34}
S. B. Ahn, D. S. Han, J. H. Bae, T. J. Byun, J. P. Kim, and C. S. Eun, “The miss rate for colorectal adenoma determined by quality-adjusted, back-to-back colonoscopies,” Gut Liver, vol. 6, no. 1, pp. 64–70, 2012, doi: 10.5009/gnl.2012.6.1.64.

\bibitem{author35}
J. Lee et al., “Risk factors of missed colorectal lesions after colonoscopy,” 2017, doi: 10.1097/MD.0000000000007468.

\bibitem{author36}
L. F. Sánchez-Peralta, L. Bote-Curiel, A. Picón, F. M. Sánchez-Margallo, and J. B. Pagador, “Deep learning to find colorectal polyps in colonoscopy: A systematic literature review,” Artif Intell Med, vol. 108, no. March, p. 101923, 2020, doi: 10.1016/j.artmed.2020.101923.

\bibitem{author37}
J. Ribeiro, S. Nóbrega, and A. Cunha, “Polyps Detection in Colonoscopies,” Procedia Comput Sci, vol. 196, pp. 477–484, 2021, doi: 10.1016/j.procs.2021.12.039.

\bibitem{author38}
N. Rani, R. Verma, and A. Jindal, “Polyp Detection Using Deep Neural Networks,” Handbook of Intelligent Computing and Optimization for Sustainable Development, pp. 801–814, Feb. 2022, doi:10.1002/9781119792642.CH37.

\bibitem{author39}
P. Rasouli et al., “The role of artificial intelligence in colon polyps detection,” Gastroenterol Hepatol Bed Bench, vol. 13, no. 3, pp. 191–199, 2020, doi: 10.22037/ghfbb.v13i3.1866.

\bibitem{author40}
F. Tang, Q. Huang, J. Wang, X. Hou, J. Su, and J. Liu, “DuAT: Dual-Aggregation Transformer Network for Medical Image Segmentation,” Dec. 2022, doi: 10.48550/arxiv.2212.11677.

\bibitem{author41}
J. Lewis, Y. J. Cha, and J. Kim, “Dual encoder–decoder-based deep polyp segmentation network for colonoscopy images,” Sci Rep, vol. 13, no. 1, pp. 1–12, 2023, doi: 10.1038/s41598-023-28530-2.

\bibitem{author42}
A. K. Mohammed, S. Yildirim-Yayilgan, I. Farup, M. Pedersen, and O. Hovde, “Y-Net: A deep Convolutional Neural Network to Polyp Detection,” British Machine Vision Conference 2018, BMVC 2018, pp. 1–11, 2019.

\bibitem{author43}
P. Song, J. Li, and H. Fan, “Attention based multi-scale parallel network for polyp segmentation,” Comput Biol Med, vol. 146, no. March, p. 105476, 2022, doi: 10.1016/j.compbiomed.2022.105476.

\bibitem{author44}
Y. Qin, H. Xia, and S. Song, “RT-Net: Region-Enhanced Attention Transformer Network for Polyp Segmentation,” Neural Process Lett, 2023, doi: 10.1007/s11063-023-11405-y.

\bibitem{author45}
H. Wu, Z. Zhao, and Z. Wang, “META-Unet: Multi-Scale Efficient Transformer Attention Unet for Fast and High-Accuracy Polyp Segmentation,” IEEE Transactions on Automation Science and Engineering, vol. PP, pp. 1–12, 2023, doi: 10.1109/TASE.2023.3292373.

\bibitem{author46}
K. Wang, Z. Qian, W. Zhang, M. Zhang, and Q. Luo, “A Novel Neural Network Based on Transformer for Polyp Image Segmentation,” 2023 IEEE 3rd International Conference on Electronic Technology, Communication and Information, ICETCI 2023, pp. 413–417, 2023, doi: 10.1109/ICETCI57876.2023.10176365.

\bibitem{author47}
Y. Wang, Z. Deng, S. Member, S. Hu, and S. Wang, “Cooperation Learning Enhanced Colonic Polyp Segmentation Based on Transformer- CNN Fusion”.

\bibitem{author48}
F. Chen, H. Ma, and W. Zhang, “SegT: Separated edge-guidance transformer network for polyp segmentation,” Mathematical Biosciences and Engineering, vol. 20, no. 10, pp. 17803–17821, 2023, doi: 10.3934/mbe.2023791.

\bibitem{author49}
W. Wang et al., “Pyramid Vision Transformer: A Versatile Backbone for Dense Prediction without Convolutions,” Proceedings of the IEEE International Conference on Computer Vision, pp. 548–558, 2021, doi: 10.1109/ICCV48922.2021.00061.

\bibitem{author50}
M. Zhu and K. Han, Vision Transformer Pruning, vol. 1, no. 1. Association for Computing Machinery.

\bibitem{author51}
Q. Chang, D. Ahmad, J. Toth, R. Bascom, and W. E. Higgins, “ESFPNet: efficient deep learning architecture for real-time lesion segmentation in autofluorescence bronchoscopic video,” 2022, Accessed: Feb. 02, 2023. [Online]. Available: http://mipl.ee.psu.edu/

\bibitem{author52}
T. C. Nguyen, T. P. Nguyen, G. H. Diep, A. H. Tran-Dinh, T. V. Nguyen, and M. T. Tran, CCBANet: Cascading Context and Balancing Attention for Polyp Segmentation, vol. 12901 LNCS. Springer International Publishing, 2021. doi: 10.1007/978-3-030-87193-2

\bibitem{author53}
R. G. Dumitru, D. Peteleaza, and C. Craciun, “Using DUCK-Net for polyp image segmentation,” Sci Rep, vol. 13, no. 1, pp. 1–12, 2023, doi: 10.1038/s41598-023-36940-5.

\bibitem{author54}
T. P. Van and S. D. Viet, ESSL-Polyp: A Robust Framework of Ensemble Semi-supervised Learning in Polyp Segmentation, vol. 739 LNNS. Springer Nature Switzerland, 2023. doi: 10.1007/978-3-031-37963-5

\bibitem{author55}
G. Liu, S. Yao, D. Liu, B. Chang, Z. Chen, and J. Wang, “CAFE-Net: Cross-Attention and Feature Exploration Network for polyp segmentation,” vol. 238, no. September 2023, 2024.

\bibitem{author56}
W. Song and H. Yu, “Non-pooling Network for medical image segmentation,” 2023, [Online]. Available: http://arxiv.org/abs/2302.10412

\bibitem{author57}
G. Liu, Y. Jiang, D. Liu, B. Chang, L. Ru, and M. Li, “A coarse-to-fine segmentation frame for polyp segmentation via deep and classification features,” Expert Syst Appl, vol. 214, no. May 2022, p. 118975, 2023, doi: 10.1016/j.eswa.2022.118975.

\bibitem{author58}
D. Jha et al., “A Comprehensive Study on Colorectal Polyp Segmentation with ResUNet++, Conditional Random Field and Test-Time Augmentation,” IEEE J Biomed Health Inform, vol. 25, no. 6, pp. 2029–2040, 2021, doi: 10.1109/JBHI.2021.3049304.

\bibitem{author59}
B. Dong, W. Wang, D.-P. Fan, J. Li, H. Fu, and L. Shao, “Polyp-PVT: Polyp Segmentation with Pyramid Vision Transformers,” 2021, [Online]. Available: http://arxiv.org/abs/2108.06932

\bibitem{author60}
F. Liu, Z. Hua, J. Li, and L. Fan, “MFBGR: Multi-scale feature boundary graph reasoning network for polyp segmentation,” Eng Appl Artif Intell, vol. 123, no. February, p. 106213, 2023, doi: 10.1016/j.engappai.2023.106213.

\bibitem{author61}
H. Al Jowair, M. Alsulaiman, and G. Muhammad, “Multi parallel U-net encoder network for effective polyp image segmentation,” Image Vis Comput, vol. 137, no. May, p. 104767, 2023, doi: 10.1016/j.imavis.2023.104767.

\bibitem{author62}
Y. Li, M. Hu, and X. Yang, “Polyp-SAM: Transfer SAM for Polyp Segmentation,” 2023, [Online]. Available: http://arxiv.org/abs/2305.00293

\bibitem{author63}
M. J. Alam and S. A. Fattah, “SR-AttNet: An Interpretable Stretch–Relax Attention based Deep Neural Network for Polyp Segmentation in Colonoscopy Images,” Comput Biol Med, vol. 160, no. September 2022, p. 106945, 2023, doi: 10.1016/j.compbiomed.2023.106945.

\bibitem{author64}
Q.-H. Trinh, “Meta-Polyp: a baseline for efficient Polyp segmentation,” 2023, [Online]. Available: http://arxiv.org/abs/2305.07848

\bibitem{author65}
M. Naderi, M. Givkashi, F. Piri, N. Karimi, and S. Samavi, “Focal-UNet: UNet-like Focal Modulation for Medical Image Segmentation,” pp. 1–8, 2022, [Online]. Available: https://arxiv.org/abs/2212.09263v1

\bibitem{author66}
N. H. Thuan, N. T. Oanh, N. T. Thuy, S. Perry, and D. V. Sang, “RaBiT: An Efficient Transformer using Bidirectional Feature Pyramid Network with Reverse Attention for Colon Polyp Segmentation,” pp. 1–15, 2023.

\bibitem{author67}
N. K. Tomar, D. Jha, and U. Bagci, DilatedSegNet: A Deep Dilated Segmentation Network for Polyp Segmentation, vol. 13833 LNCS. Springer International Publishing, 2023. doi: 10.1007/978-3-031-27077-2

\bibitem{author68}
N. K. Tomar, D. Jha, U. Bagci, and S. Ali, “TGANet: Text-Guided Attention for Improved Polyp Segmentation,” in Lecture Notes in Computer Science (including subseries Lecture Notes in Artificial Intelligence and Lecture Notes in Bioinformatics), 2022, pp. 151–160. doi: 10.1007/978-3-031-16437-8

\bibitem{author69}
P. Ngoc Lan et al., “NeoUNet: Towards Accurate Colon Polyp Segmentation and Neoplasm Detection,” Lecture Notes in Computer Science (including subseries Lecture Notes in Artificial Intelligence and Lecture Notes in Bioinformatics), vol. 13018 LNCS, pp. 15–28, 2021, doi: 10.1007/978-3-030-90436-4

\bibitem{author70}
T. Yu and Q. Wu, “HarDNet-CPS: Colorectal polyp segmentation based on Harmonic Densely United Network,” Biomed Signal Process Control, vol. 85, no. June 2022, p. 104953, 2023, doi: 10.1016/j.bspc.2023.104953.

\bibitem{author71}
R. Karmakar and S. Nooshabadi, “Mobile-PolypNet: Lightweight Colon Polyp Segmentation Network for Low-Resource Settings,” J Imaging, vol. 8, no. 6, 2022, doi: 10.3390/jimaging8060169.

\bibitem{author72}
T. P. Van, L. B. Doan, T. T. Nguyen, D. T. Tran, Q. Van Nguyen, and D. V. Sang, “Online pseudo labeling for polyp segmentation with momentum networks,” Proceedings - International Conference on Knowledge and Systems Engineering, KSE, vol. 2022-Octob, pp. 1–6, 2022, doi: 10.1109/KSE56063.2022.9953785.

\bibitem{author73}
J. Silva, A. Histace, O. Romain, X. Dray, and B. Granado, “Toward embedded detection of polyps in WCE images for early diagnosis of colorectal cancer,” Int J Comput Assist Radiol Surg, vol. 9, no. 2, pp. 283–293, 2014, doi: 10.1007/s11548-013-0926-3.

\bibitem{author74}
J. Bernal, F. J. Sánchez, G. Fernández-Esparrach, D. Gil, C. Rodríguez, and F. Vilariño, “WM-DOVA maps for accurate polyp highlighting in colonoscopy: Validation vs. saliency maps from physicians,” Computerized Medical Imaging and Graphics, vol. 43, pp. 99–111, Jul. 2015, doi: 10.1016/J.COMPMEDIMAG.2015.02.007.

\bibitem{author75}
D. Vázquez et al., “A Benchmark for Endoluminal Scene Segmentation of Colonoscopy Images,” J Healthc Eng, vol. 2017, 2017, doi: 10.1155/2017/4037190.

\bibitem{author76}
P. Ngoc Lan et al., “NeoUNet: Towards Accurate Colon Polyp Segmentation and Neoplasm Detection,” Lecture Notes in Computer Science (including subseries Lecture Notes in Artificial Intelligence and Lecture Notes in Bioinformatics), vol. 13018

\end{thebibliography}

\end{document}